\documentclass[10pt,a4]{article}

\usepackage{graphicx}
\usepackage{dcolumn}
\usepackage{bm}
\usepackage{amsmath}
\usepackage{amsfonts}
\usepackage{amssymb}

\begin{document}


\title{Non-linear density functional theory: A direct method to
calculate many-electron charge densities.}

\author{Werner A. Hofer and Kriszti\'{a}n Palot\'{a}s \\ {\small
         Surface Science Research Centre,
         University of Liverpool, Liverpool L69 3BX, Britain}\\
         {\small Donostia International Physics Center, San Sebastian,
         Spain}}
\maketitle
\begin{abstract}
We suggest to include the density of electron charge explicitly in
the electron potential of density functional theory, rather than
implicitly via exchange-correlation functionals. The advantages of
the approach are conceptual and numerical. Conceptually, it allows
to formulate a physical principle for the development of quantum
mechanical systems: it is the principle of energetic equilibrium,
because the energy principle, in this case, applies not only
globally, but also on a local level. The method is an order-$N$
method, scaling linearly with the number of atoms. It is used to
calculate the electronic groundstate of a metallic surface, where
we find good agreement with experimental values.
\end{abstract}



\section{Density functional theory}
A key innovation in theoretical solid state physics in the last
fifty years was the reformulation of quantum mechanics in a density
formalism, based on the Hohenberg-Kohn theorem \cite{hk64}. Despite
initial resistance, in particular from quantum chemists, the method
has replaced previous frameworks and provides, to date, the most
advanced theoretical model for the calculation of atoms, solids, and
liquids. However, its implementation relies on a rather cumbersome
detour. While the Hohenberg-Kohn theorem is formulated exclusively
in terms of electron charge densities and energy functionals,
calculations today are based almost exclusively on the
specifications given by Kohn and Sham one year after the initial
theorem was made public \cite {ks65}. These Kohn-Sham equations
define, how the electron interactions within a system can be split
into three separate single-electron parts: a Coulomb potential
$U_{Cou}$, giving the electron-ion attraction and the
electron-electron repulsion; an exchange potential $U_X$, taking
into account the Pauli exclusion principle; and a correlation
potential $U_C$, covering the correlated motion of electrons in a
many-electron system. Then every electron in the system is described
by a single-particle Schr\"odinger equation with the effective
potential $U_{eff}$:
\begin{equation}
U_{eff} = U_{Cou} + U_X + U_C
\end{equation}
in order to determine the charge density. Within the local density
approximation, the exchange and correlation potentials are usually
combined and thought to be depending on the local density of
electron charge $\rho({\bf r})$:
\begin{equation}
U_{XC}({\bf r}) = U_{XC}(\rho({\bf r}))
\end{equation}
The variational problem thus contained at every single step of the
energy minimization cycle a solution of the single-particle
Schr\"odinger equation, from which the electron eigenstates and
their eigenvalues are determined. The density of charge at a given
step of the cycle is calculated by adding single-electron charges:
\begin{equation}
\rho({\bf r}) = \sum_{i=1}^N \psi_i({\bf r})\psi_i^*({\bf r})
\end{equation}
$N$ denotes the total number of electrons in the system. While this
procedure is generally successful, and implemented today in
numerical methods optimized for efficiency (see e.g the ingenious
way ionic and electron degrees of freedom are treated on much the
same footing following a method developed by Car and Parinello
\cite{cp85}), it is highly inefficient in one crucial conceptual
point: If, according to the Hohenberg-Kohn theorem, the density of
charge is the only physically relevant variable of the system, then
solving the Schr\"odinger equation, setting up the eigenvectors, and
computing the density of electron charge is an operation, which
creates a vast amount of redundant information. Every information,
pertaining to the solution of the single-particle Schr\"odinger
equation and the summation of single electron charges is discarded
at the end of every step in the iteration cycle. One could therefore
say that more than 90\% of the information created in today's
simulations is actually irrelevant. The question thus arises: {\em
Do we have to create this information at all, or can we find a more
direct way to arrive at the groundstate density of charge without
this cumbersome detour via the single-particle Schr\"odinger
equation?}

From a physical point of view, this state of affairs is easy to
understand, if one considers the basic innovation, quantum mechanics
introduced into the original framework of mechanics. It is the
property of {\em phases} and {\em phase correlations}, which, even
though it is present in any many-electron system, does not find an
expression in its key quantity, the density of charge. Inducing
phase correlations into density functional theory thus meant to
rebuild a phase coherent system by way of the original Schr\"odinger
equation. The easiest way to build such a system is to determine the
coherent wavefunctions of single electrons and to ensure that the
states are orthogonal in Hilbert space.

A more general approach would be to look at phase-correlations
within a different framework. This framework, the theory of
density matrices \cite{hg95,wl95,bhg97}, has consequently been the
focus of research over a period of more than ten years. However, a
framework combining the notion of phase-coherence for
many-electron wavefunctions, and the basic requirement in DFT to
find the groundstate density, has so far been missing. In this
paper, we introduce such a {\em direct} method and show how it can
be used to determine the groundstate density without any recourse to
the single-particle Schr\"odinger equation.

The paper was written in the spirit of {\em orbital free} density
functional theory \cite{levy84,dreizler90,wang99}. It is frequently
pointed out by protagonists of this line of research, that an
orbital free, and Bloch-state free formulation of DFT has enormous
potentials for numerical improvements. The main difference to the
present framework is philosophical, rather than practical. While
orbital free developments are essentially based on the
Hohenberg-Kohn theorem and attempt to find a transferable
formulation for the kinetic energy functional, we have analyzed the
foundation of the theory itself and sought a continuous
reformulation of the fundamental theory. Not least, because such a
reformulation has substantial conceptual advantages, as will be
shown in the following sections.

The paper is organised as follows: in section II we introduce the
theoretical concept, based on a continuous generalisation of
potentials in the many-electron Schr\"odinger equation. It is
shown that the ensuing equation for the density of charge is
non-linear, and that Sch\"odinger's equation is a zero-order
approximation of the general relation. In section III we sketch
the solution of the eigenvalue problem for a solid-state system.
Finally, the new formulation is tested for metal surfaces in
section IV.

\section{Theoretical basis}

The theorem of Hohenberg and Kohn elevated the density of charge
to the state of a {\em physical} variable in many-electron
systems. In addition, observations by scanning tunnelling
microscopes seem to confirm that the observed structures at a
surface are essentially due to the distribution of electron
charge, measured with a resolution of a few picometer
\cite{hofer03}. It is then quite astonishing to find that the
electron charge has to be determined by a complicated computation
process from the single electron states. We show in the following,
how a representation of the system, without any reference to these
states, can be constructed on the basis of Green's functions.

\subsection{Green's functions and phase correlations}

The Green's function of a many-electron system can be expanded into
single electron states by a spectral decomposition
\cite{palotas05,datta95}:
\begin{equation}\label{000}
G({\bf r},{\bf r}', z) = \sum_{i=1}^N \frac{\psi_i({\bf
r})\psi_i^*({\bf r}')}{z - \epsilon_i}
\end{equation}

Here, $G({\bf r},{\bf r}', z)$ is the many-electron Green's
function of the system in a local basis, $N$ the number of
electrons, and $z$ the complex energy parameter. The Green's
function will become its complex conjugate upon an exchange of the
local coordinates:
\begin{equation}\label{001}
G({\bf r},{\bf r}', z) = G^*({\bf r}',{\bf r}, z^*)
\end{equation}
It complies with a differential equation of the Hamiltonian in a
local basis:
\begin{equation}\label{002}
\left(H({\bf r},{\bf r}') - z \right) G({\bf r},{\bf r}', z) =
-\delta({\bf r} - {\bf r}')
\end{equation}
Note that the Green's function of the many-electron system depends
only on two local coordinates ${\bf r}$ and ${\bf r}'$. We show now
that this Green's function can be written as the product of the
phase correlation functions $\Psi({\bf r})$ and $\Psi({\bf r}')$ and
an energy function $F(z)$:
\begin{equation}\label{003}
G({\bf r},{\bf r}',z) = F(z) \Psi({\bf r})\Psi^*({\bf r}')
\end{equation}
where $|\Psi({\bf r})|^2$ is equal to the density of charge upon a
suitable choice of $F(z)$. First, it follows from Eq. \ref{003},
that the conjugate complex of the $F(z) \Psi({\bf r})\Psi^*({\bf
r}')$ complies with the same relation as the Green's function, i.e.
it is equal to the transposed Green's function:
\begin{equation}
G^*({\bf r}',{\bf r}, z^*) = \left[F(z^*) \Psi({\bf
r}')\Psi^*({\bf r})\right]^* = F(z) \Psi^*({\bf r}')\Psi({\bf r})
= G({\bf r},{\bf r}', z)
\end{equation}
Taking the diagonal Green's function in the limit of real
eigenvalues gives:
\begin{equation}
G^{\pm}({\bf r},{\bf r},E)=\Psi({\bf r})\Psi^*({\bf r})
\lim_{\eta\rightarrow +0}F(E\pm i\eta)
\end{equation}
The imaginary part of the diagonal Green's function, integrated
over energy, must be equal to the charge density:
\begin{eqnarray}
\rho({\bf r})&=&\mp\frac{1}{\pi} \int_{-\infty}^{+\infty}dE Im
G^{\pm}({\bf r},{\bf r},E)\nonumber \\ &=&\mp\frac{1}{\pi}
\int_{-\infty}^{+\infty}dE Im \lim_{\eta\rightarrow +0}F(E\pm i\eta)
\Psi({\bf r})\Psi^*({\bf r})
\end{eqnarray}
If $\Psi({\bf r})\Psi^*({\bf r})=\rho({\bf r})$, the rest on the
right side must be unity. This is satisfied if
\begin{equation}
\mp\frac{1}{\pi} Im \lim_{\eta\rightarrow +0}F(E\pm
i\eta)=\delta(E-E_0)
\end{equation}
Using the fact that
\begin{equation}
\lim_{y\rightarrow +0}\frac{1}{x\pm iy}=
\mathbb{P}\left(\frac{1}{x}\right)\mp i\pi\delta(x)
\end{equation}
gives us the explicit form of $F(z)$:
\begin{equation}
F(z)=\frac{1}{z-E_0}
\end{equation}
Thus, the Green's function can be written as the product of two
local phase correlation functions\cite{datta95}:
\begin{equation}
G({\bf r},{\bf r}',z)=\frac{\Psi({\bf r})\Psi^*({\bf r}')}{z-E_0}
\end{equation}

At this stage we have shown that the phase correlation function
 $\Psi({\bf r})$ is an {\em exact} representation of the $N$-electron system,
 as long as the system can be described by a Green's function
 Eq. \ref{000}. However, it is not yet
clear, whether there is a differential equation describing the
evolution of $\Psi({\bf r})$. This equation cannot be equal to the
standard Kohn-Sham equations, as this would lead to a non-local
concept. If we write the Hamiltonian $H({\bf r},{\bf r}')$ as the
sum:
\begin{equation}
H({\bf r},{\bf r}') = H({\bf r}) + H({\bf r}') - H_1({\bf
r})H_1^*({\bf r}'),
\end{equation}
and insert into Eq. \ref{002} we get:
\begin{eqnarray}
& &\frac{\Psi^*({\bf r}') \left[H({\bf r}) - z \right] \Psi({\bf
r})}{z - E_0} + \frac{\Psi({\bf r}) \left[H({\bf r}') - z \right]
\Psi^*({\bf r}')}{z - E_0} - \\ &-& \frac{\left[H_1({\bf
r})H_1^*({\bf r}') - z \right] \Psi({\bf r})\Psi^*({\bf r}')}{z -
E_0} = -\delta({\bf r} - {\bf r}')\nonumber
\end{eqnarray}

It can be seen that the second line contains a product of operators
at different locations ${\bf r}$ and ${\bf r}'$. This would mean
that the phase correlation at a point ${\bf r}$ depends on events at
point ${\bf r}'$, even though there is no obvious connection.
However, since $\Psi$ depends only on ${\bf r}$, this seems
unjustified. In addition, the Hohenberg-Kohn theorem proves that if
$|\Psi({\bf r})|^2$ is equal to the
groundstate density of charge, then an energy eigenvalue $E_0$ is
defined and must be equal to the total energy. Therefore it should
be possible to formulate the problem of finding the phase
correlation function $\Psi({\bf r})$ in an eigenvalue equation. In
this case the many-electron Green's function has only one pole, in
contrast to the standard spectral decomposition, where it has $N$
poles.

Here we make the only conjecture in this paper: we assume that the
phase correlation function $\Psi({\bf r})$ is described by a
suitable modification of a single coordinate Schr\"odinger equation:
\begin{equation}
- \frac{\hbar^2}{2m} \nabla^2 \Psi({\bf r}) + U({\bf r})\Psi({\bf
r}) = E_0 \Psi({\bf r})
\end{equation}
with a local potential $U({\bf r})$. That such a modification is
possible and leads to correct results in simple test cases, will
be shown in the following. The conjecture is supported by the
derivation of an approximate single coordinate Schr\"odinger
equation from the Hartree-Fock model of an $N$-electron system
\cite{bao03}.

If this conjecture is justified, then we have obtained, at this
point, a phase correlation function, which includes all phase
correlations in the occupied range, it complies with a local
equation, its corresponding eigenvalue $E_0$ is equal to the total
energy of the system, and its square gives the electron charge
density in the system. This entity looks very much like a descriptor
of the physical system itself. Something, which is otherwise called
a many-body wavefunction. As this term might lead to confusion,
since the many-body wavefunction is in standard theory described by
$N$ local coordinates (${\bf r}_1,{\bf r}_2,{\bf r}_3, ...,{\bf
r}_N$), we shall call it the {\em many-electron} wavefunction in the
rest of the paper.

An additional note concerning the relation of the framework
presented in this paper to density functional theory seems
necessary. In particular, since the original derivation of the
Hohenberg-Kohn theorem is based on many-body wavefunctions
\cite{hk64}. In this respect we note that the groundstate charge
density, once obtained, can be written in any representation. Thus
an energy minimum, associated with a density, is enough to
describe the physical content of a many-electron system. As long
as the energy is a minimum, it is therefore irrelevant, how this
minimum was obtained in practice. This is reflected by the most
efficient methods in density functional theory, which are based on
trial wavefunctions, which do not diagonalize the Hamiltonian.
Instead, the off-diagonal elements are minimized with the help of
numerical algorithms until they vanish.

\subsection{The non-linear Schr\"odinger equation}

So far, we have not considered the modified form of the
Hamiltonian $H({\bf r})$. In particular the potential, which in
the standard formulation is the potential of single electrons, is
not suitable for our purposes. To find a local formulation of the
problem, we start with the Schr\"odinger equation of the
many-electron system, written as:
\begin{equation}\label{SE}
- \frac{\hbar^2}{2m} \nabla^2 \Psi({\bf r}) + U({\bf r})\Psi({\bf
r}) = E \Psi({\bf r})
\end{equation}
Here, $\nabla^2$ denotes the Laplacian, $U({\bf r})$ a potential
in the system, and $E$ the total energy. Within the context of
density functional theory it is generally assumed that the
potential $U$ reflects the interaction between the charge of one
electron, and the potential of the environment. This entails that
the potential is a sum containing the positions of all electrons
in the system. While such an approach is possible, it does not
reflect the previous findings. If phase correlations and charge
are local quantities, then the interactions between electron
charge and the potential of all other charges should also be
local.

The single coordinate equation \ref{SE} must therefore be recast
into a relation based on densities, and not integral quantities.
This provides the desired equation for the many-electron
wavefunction. If $U({\bf r})$ is e.g. the Coulomb potential of an
ion with charge $Z$, then the potential density of electron
attraction at a point $({\bf r})$ is given by:
\begin{equation}
v({\bf r}) \rho({\bf r}) = - \frac{1}{4\pi \varepsilon_0}
\frac{Ze^2 \rho({\bf r})}{|{\bf r} -
{\bf R}_{ion}|} \qquad v({\bf r}) = - \frac{1}{4\pi \varepsilon_0}
\frac{Ze^2}{|{\bf r} - {\bf R}_{ion}|}
\end{equation}
Similarly, the kinetic energy term and the total energy will
change into kinetic energy density and total energy density. The
variables in the local equation will be changed into:
\begin{equation}
\frac{\hbar^2}{2m} \Rightarrow 2 f \qquad U({\bf r}) \Rightarrow
v({\bf r}) \rho({\bf r}) \qquad E \Rightarrow \epsilon
\end{equation}
$f$ is a system parameter, related to the volume of the system,
which shall be analysed in detail further down. $v({\bf r})$ is the
external potential. $\epsilon$ is the energy density. The equation
then reads:
\begin{equation}\label{CSE}
- 2f \nabla^2 \Psi({\bf r}) + v({\bf r}) \rho({\bf r}) \Psi({\bf
r}) = \epsilon \Psi({\bf r})
\end{equation}

\subsubsection{External potential}
The interaction between electrons and ions in a groundstate
many-body problem is exclusively described by Coulomb
interactions. If $\Psi({\bf r})$, as shown in previous sections,
uniquely describes the problem, then the same should hold for the
modified Schr\"odinger equation. Exchange correlation potentials,
as in density functional theory, would enter the description only
at the level of single electron states. Since $\Psi({\bf r})$
describes the total system, and not single electron state, they
will not enter the present framework. However, Coulomb
interactions will be modified, if one considers magnetic systems.
In this case the potential $v({\bf r})$ will depend on the
spin-state of electron charge. At present, we do not consider this
case. The potential $v({\bf r})$ then is the Coulomb interaction
potential due to electrons and ions,
\begin{equation}
v({\bf r})=\frac{e^2}{4\pi\varepsilon_0}\left[
\int d^3r'\frac{\rho({\bf r}')}{|{\bf r} - {\bf r}'|}-
\sum_{i=1}^{M}\frac{Z_i}{|{\bf r} - {\bf R}_i|}\right]
\end{equation}

\subsubsection{Charge density}
The density of charge $\rho({\bf r})$ is the square of the
many-electron wavefunction, or:
\begin{equation}
\rho({\bf r}) = \Psi^*({\bf r}) \Psi({\bf r})
\end{equation}
Writing an equivalent equation for the complex conjugate function
$\Psi^*({\bf r})$, we get:
\begin{equation}
- 2f \nabla^2 \Psi^*({\bf r}) + v({\bf r}) \rho({\bf r})
\Psi^*({\bf r}) = \epsilon \Psi^*({\bf r})
\end{equation}
Multiplying the first of these equations by $\Psi^*({\bf r})$, and
the second by $\Psi({\bf r})$, and adding the equations, we end up
with:
\begin{equation}\label{01}
- f \left[\Psi^*({\bf r}) \nabla^2 \Psi({\bf r}) + \Psi({\bf
r})\nabla^2\Psi^*({\bf r})\right] + v({\bf r}) \rho({\bf r})
\rho({\bf r}) = \epsilon \rho({\bf r})
\end{equation}
The Laplacian acting on the density is given by:
\begin{eqnarray}
&\nabla^2& \left[\Psi({\bf r})\Psi^*({\bf r})\right] =
\left[\Psi^*({\bf r}) \nabla^2 \Psi({\bf r}) + \Psi({\bf
r})\nabla^2\Psi^*({\bf r})\right] + \nonumber \\ &+& 2 \left[\nabla
\Psi({\bf r})\right] \left[\nabla \Psi^*({\bf r})\right]
\end{eqnarray}
The second term on the right contains the essential phase
correlations of the many-electron wavefunction. We symbolize it by
$\Pi[\Psi]$ to denote that it contains the many-electron phase
correlations. In principle there seems no straightforward method
to describe it in terms of charge densities alone: as the
many-electron system has to be phase coherent, the density of
charge alone provides insufficient information. If this were not
the case, then a formulation based on density alone would be
sufficient to describe the physical content of a many-electron
system. From this viewpoint, the essential finding of density
functional theory is not only the uniqueness of the charge
distribution, but also of the groundstate energy. This energy, in
turn, can only be determined by constructing a phase coherent
system. Thus the density of charge will be described by:
\begin{eqnarray}
\left[ -f \nabla^2 + v({\bf r}) \rho({\bf r}) \right]\rho({\bf r})
 + f \Pi[\Psi] & = & \epsilon \rho({\bf r}) \nonumber \\
 \Pi[\Psi] & = & 2 \left[\nabla
\Psi({\bf r})\right] \left[\nabla \Psi^*({\bf r})\right]
\end{eqnarray}
However, the non-linear equation describes the
 many-electron wavefunction, provided the potential and
groundstate charge density are known. It is therefore not strictly
necessary to find a formulation based on density alone, as long as
the nonlinear equation leads to correct many-electron
wavefunctions. The derivation thus provides a set of three
equations, defining the groundstate of a system:
\begin{eqnarray}\label{set}
\left[- 2f \nabla^2  + v({\bf r}) \rho({\bf r})\right] \Psi({\bf
r}) & = & \epsilon \Psi({\bf r}) \nonumber \\
\left[ -f \nabla^2 + v({\bf r}) \rho({\bf r}) \right]\rho({\bf r})
 + f \Pi[\Psi] & = & \epsilon \rho({\bf r}) \\
 \rho({\bf r}) & = & \Psi({\bf r}) \Psi^*({\bf r}) \nonumber
\end{eqnarray}

 Note that this set of equations applies to the many-electron density of charge without any
reference to single Kohn-Sham states. It describes the physical
system by four variables: the density of charge $\rho({\bf r})$,
the external potential $v({\bf r})$, which will also depend on the
density, the energy density $\epsilon$, and the many-electron
wavefunction $\Psi({\bf r})$. The equations lead to a direct
method of calculating the groundstate charge density by solving
the nonlinear Schr\"odinger equation and minimizing the energy
density $\epsilon$ of the system. The self-consistent procedure,
elaborated in the following section, will be
\begin{itemize}
\item make an initial guess for the density $\rho_0$, \item
construct the potential $v_0$, \item solve the nonlinear
Schr\"odinger equation for $\Psi$, and \item construct a new
charge density $\rho_1=\Psi^* \Psi$.
\end{itemize}
The cycle is repeated until input and output charge densities are
equal. The problem of finding the groundstate energy density can
be formulated as the following minimization problem:
\begin{equation}
\epsilon = \left\{\frac{\int d^3{\bf r} \left(\left[- f \nabla^2 +
v({\bf r}) \rho({\bf r})\right]\rho({\bf r}) + f \Pi[\Psi({\bf
r})]\right)}{\int d^3{\bf r} \rho({\bf r})}\right\}_{Min}
\end{equation}
Note that also in this case the problem cannot be formulated as a
minimization problem of charge density alone. This is a
consequence of basing the whole derivation on the many-electron
problem and ensuring the phase coherence throughout the system.

\subsection{Physical implications}

From a dimensional point of view it is interesting to analyse the
meaning of the system constant $f$. Its dimension is that of a
force:
\begin{equation}
\left[f \nabla^2\right] = \left[\frac{E}{L^3}\right] \qquad [f] =
\left[\frac{E}{L^3} L^2\right] = \left[\frac{E}{L}\right] = N(SI)
\end{equation}
Apart from the universal constants $\hbar$ and $m$ the system
constant $f$ only depends on the volume. In this respect, it is
beyond the well known Thomas-Fermi model, where the kinetic energy
density is system independent. The second difference is that phase
correlations are included in the picture, by solving the
non-linear Schr\"odinger equation. The differences are crucial. As
shown in the following sections, the model recaptures the density
oscillations at metal surfaces, contrary to the Thomas-Fermi
model. That the kinetic energy density {\em cannot} be system
independent, is in line with observations in quantum mechanics
that the energy of a system comprising e.g. a single electron is
inverse proportional to the enclosing volume. It seems therefore
that the action of an external potential on the density of charge
can be seen as a force acting on the wavevector, or the spatial
distribution of the density. The reason for this force is that
energy at a given point of the system has to be conserved. The
reason that it acts on the curvature rather than the amplitude is
the relation between curvature or wavevector and kinetic energy,
predicted by Louis de Broglie, and experimentally confirmed by
Davisson and Germer \cite{ldb25,dg27}.

Since $\epsilon$ was defined as the total energy divided by the
volume the equation has an additional physical content. A solution
of the equation with constant $\epsilon$ means that the energy
density does not vary from one point of the system to another.
From the viewpoint of thermodynamics, this is also the condition
of a system in equilibrium. One could thus say that the specific
form of the Schr\"odinger equation has one very specific physical
meaning: it provides the solution of distributing electron charge
in an external potential in such a way, that the energy density is
constant. It explains, why the mathematical form of the equation
is different from e.g. a differential equation in classical
electrodynamics. There, the boundary conditions usually lead to a
selection of amplitudes and frequencies in a system. Here, the
frequencies need to be modified at every single point in order to
achieve constant energy density values. In principle, this is
still a wave-dynamical problem, but a problem modified by the
ability of electrons to interact with external potentials and with
each other. This formulation of the many-electron problem also
explains, why wave mechanics, based on the Schr\"odinger equation,
{\em appears} to be non-local \cite{selleri88}: since the main
condition implicitly contained in the equation is the condition of
thermal equilibrium, one cannot separate one part of the system
from any other part, without contradicting the assumption that the
system as a whole will equilibrate. Thus, wave mechanics is in
fact a local theory, like every other theory known in physics, but
it is applied to systems which are assumed to be connected
throughout. This can clearly be seen in the mathematical form of
Eqs. \ref{set}: all variables of the system, $v({\bf
r})$,$\rho({\bf r})$, and $\epsilon$ are local scalars, defined at
a specific point of the system. These local variables cannot,
however, be arbitrarily chosen, since $\epsilon$ {\em must} be a
constant and $f$ {\em must} relate to the volume of the system.

One might object, at this point, that the problem is now vastly
more complicated. After all, the eigenvalue problem is now
non-linear, and it is generally impossible to compute solutions of
non-linear differential equations in full generality. Here, we
remember the main result of the Hohenberg-Kohn theorem: the
groundstate charge density is the charge density, which minimizes
the total energy and thus the energy density $\epsilon$. This
shall provide, as will be seen, a way to compute the exact
solution starting from an initial guess for the charge density
$\rho({\bf r})$, by solving a linear eigenvalue equation and
adjusting the charge density from one iteration to the next. In
principle, it is a very similar method to the one used today in
standard DFT codes, but for two decisive advantages:

\begin{enumerate}
\item The charge density is computed directly without any
reference to single-particle Schr\"odinger equations. \item The
charge density computed is the many-electron charge density and it
is therefore completely independent of exchange-correlation
functionals and density approximations.
\end{enumerate}

\subsection{Possible objections}

From a pragmatic point of view, the approach seems not too
different to the standard approach, apart from the fact that the
non-local density-correlations in the potential have been replaced
by the charge density itself. However, there is a number of
issues, related to non-linear concepts, which deserve a closer
look.
\begin{itemize}
\item If the Hohenberg-Kohn theorem is correct, then the
groundstate charge density is unique. Here, we have a non-linear
potential, and the groundstate is therefore potentially not
unique.
\end{itemize}
There are two possible answers. The first is the current state of
DFT itself. After all, exchange correlation functionals {\em
include} the density and its derivatives. The effective potential in
today's DFT simulations is therefore also non-linear. This, it
seems, does not invalidate the results of calculations. The second
is based on recent research (in Mathematics) on the behavior of
non-linear Schr\"odinger equations, related to weak coupling in a
many-electron systems \cite{bao03}: in this case the equations still
yield physically acceptable solutions, even though the existence of
a solution depends on reasonable physical conditions.
\begin{itemize}
\item The Thomas-Fermi model also attempts to calculate charge
densities from a non-linear potential and a kinetic energy
density. The same seems to be tried here. But we know that this
model fails even to explain chemical bonding.
\end{itemize}
The answer to this question lies in the treatment of the kinetic
energy term, and the use of the nonlinear Schr\"odinger equation to
determine many-electron charge densities. Within the Thomas-Fermi
model, the prefactor of the kinetic energy functional is a constant.
This means, that it does not depend on the system, but is
universally the same. Physically speaking, this cannot be correct.
Because an electron in a very large negative potential {\em must}
experience different kinetic effects than an electron in a
comparatively small negative potential. In the present model the
kinetic energy density is inversely proportional to the volume, and
contains a term which is directly related to the many-electron phase
correlations. It will be seen in the simulation of metal surfaces,
that this leads to the correct behavior of electrons at a boundary.
\begin{itemize}
\item I do not see spin-statistics involved in this approach. In
standard DFT this is part of the exchange-correlation functionals.
Where is it in this formulation?
\end{itemize}
Here, one has to note that spin-statistics are also not involved
in the formulation of the non-magnetic  many-electron
Schr\"odinger equation. They only arise at the level, where single
electrons are thought of as individual entities. A many-electron
wavefunction in a many-electron Schr\"odinger equation is a unique
scalar function of location and spin. In a non-magnetic
formulation only of location. So far, we have not included spin in
the framework. This must be done in field mediated manner and will
be the topic of future work.
\begin{itemize}
\item Non-linearity is nothing new. It is already present in the
standard techniques, since self-consistency cycles are essentially
non-linear. The non-linearity lies therefore in the variational
principle, but not in the fundamental equations.
\end{itemize}
While we do not dispute that the variation and the ensuing
self-consistency cycles are non-linear, the fundamental equations,
the Kohn-Sham equations, are still linear. If this were not the
case, then the groundstate of a system {\em could not} be
described as a superposition of $N$ single electron states. While
this, we think, is correct for a mean-field theory, it is not
suitable within a many-electron framework. In this sense the
described approach extends the non-linearity to the fundamental
equations themselves, with the effect, that superpositions are no
longer possible. The groundstate calculated with this approach
cannot be decomposed into single-electron states, or only, if one
uses an approximation.

Before sketching the solution of the eigenvalue problem, let us
consider some limiting cases, where one can show that the linear
Schr\"odinger equation is in fact the zero order approximation of
the general problem.

\subsection{Zero order approximations}

The main areas, where the Schr\"odinger equation is used today,
are the theory of atoms, and solid state theory. In solid state
theory the most frequently employed DFT codes use periodic
boundary conditions and are generally applied to systems with
translational symmetry. In quantum chemistry, DFT methods usually
focus on atomic orbitals and their representation by suitable
basis sets.

Let us consider initially an atomic system, where the density of
charge can be closely mimicked by an exponential function. Thus,
with radial symmetry:
\begin{equation}
\rho(r) = \rho_0 \exp \left(-\kappa r\right)
\end{equation}
The expansion of the exponential function gives:
\begin{equation}
\rho(r) = \rho_0 \left(1 - \frac{\kappa r}{1!} + \frac{(\kappa
r)^2}{2!} - \frac{(\kappa r)^3}{3!} + ... \right)
\end{equation}
The zero order approximation assumes that $\rho(r) = \rho_0$ =
constant. Inserting $\rho_0$ into Eq. \ref{CSE}, and multiplying
by the volume $V$, we get:
\begin{equation}
- 2f V \nabla^2 \Psi({\bf r}) + v_{nl}({\bf r}) \rho_0 V \Psi({\bf
r}) = E \Psi({\bf r})
\end{equation}
Setting $2f V = \hbar^2/2m$ we arrive at the conventional
Schr\"odinger equation, e.g. for the hydrogen atom:
\begin{equation}
\left[ - \frac{\hbar^2}{2 m} \nabla^2 + U({\bf r}) \right]
\Psi({\bf r}) = E \Psi({\bf r})
\end{equation}
In solid state physics, systems are generally periodic. The density
of charge in this case will be a periodic function. For the sake of
demonstration we use a cosine, but the same argument applies to a
sine or an imaginary exponent. Say, we can write the density of
charge as:
\begin{eqnarray}
\rho({\bf r}) &=& \alpha (1 + \beta \cos \left({\bf k}{\bf
r}\right)) = \alpha \nonumber + \\ &+& \alpha \cdot \beta \left(1
- \frac{({\bf k}{\bf r})^2}{2!} + \frac{({\bf k}{\bf r})^4}{4!} -
\frac{({\bf k}{\bf r})^6}{6!} + ... \right)
\end{eqnarray}
where ${\bf k}$ is a vector in reciprocal space, and $\beta < 1$.
We set $\alpha + \alpha \cdot \beta = \rho_0$. Then the zero order
approximation for the non-linear eigenvalue problem will read (we
have again multiplied by the volume $V$):
\begin{equation}
- 2f V \nabla^2 \Psi({\bf r}) + v_{nl}({\bf r}) \rho_0 V \Psi({\bf
r}) = E \Psi({\bf r}),
\end{equation}
and it will again lead to the conventional Schr\"odinger equation
if we set $2f V = \hbar^2/2m$:
\begin{equation}
\left[ - \frac{\hbar^2}{2 m} \nabla^2 + U({\bf r}) \right]
\Psi({\bf r}) = E \Psi({\bf r})
\end{equation}
We are well aware that these examples do not constitute a proof
that the non-linear eigenvalue problem is exactly equivalent to
the Schr\"odinger equation. In fact, if this were the case, then
the whole formulation might be considered redundant, since it is
only a different way of stating what is already known.

\subsection{Removing the paradoxes from quantum mechanics}

The new framework removes two of the fundamental problems in
quantum mechanics, which have been the subject of debate almost
from the beginning of modern physics. It is probably not
exaggerated to say that the research done on quantum paradoxes
has, over the years, produced a volume of work comparable to the
volume of work in standard developments.

\begin{enumerate}
\item The Schr\"odinger cat paradox \cite{schrodinger35}. In the
standard formulation the cat is either dead, or alive, but its
state can only be determined by a measurement. There is,
fundamentally, no objective way of deciding before a measurement.

\item The Einstein-Podolsky-Rosen paradox \cite{epr35}. In this
case a quantum mechanical system is thought to be split into two
subsystems which increase their distance with the velocity of
light. After some time two separate measurements are performed on
either subsystem, and according to the standard formulation, the
measurement on one subsystem should determine the measurement on
the second one, even though no physical interaction is known which
exceeds the speed of light. \end{enumerate}

Both paradoxes rest on one fundamental feature of systems in
standard quantum mechanics: the first, on the possibility of
superimposing separate solutions of the Schr\"odinger equation to
describe the state of a system; the second, on the very same
feature with a slight modification concerning the times of
measurement and the critical distance between the subsystems. The
resolution of the paradoxes, within this new framework, is the
easiest possible answer: the cat is dead, or it is alive,
independent from any measurement, because there exists at any
given moment only one unique solution to the fundamental
equations. This is a consequence of the non-linearity of the
general framework, as already pointed out. The same applies to the
second paradox, with the slight modification, that performing a
measurement on one subsystem changes the state of this subsystem.
While in the standard framework, this has an effect on the second
subsystem, since the state of the overall system is described by a
superposition, it does not affect the second subsystem in the
present framework, since the fundamental principle of
equilibration would be violated. The second subsystem {\em cannot}
equilibrate with the first one, {\em because} it cannot interact.

From a physical point of view, this resolution of the paradoxes
seems actually more in keeping with a materialistic mindset.
Physicists of a classical disposition have been arguing against
the consequences of the mathematical framework, and in particular
the feature of superposition, for almost a century. Here, we adopt
the view that this consequence hinges on the linearity of the
Schr\"odinger equation, and that this linearity is actually a
feature enforced by the approximations used.

\section{Iterative solution of the non-linear eigenvalue problem}

While some eighty years ago the non-linear problem would have been
unsolvable, this has changed significantly over the last fifty
years due to the existence of powerful computers. In this respect,
theory is not much different from experiments: a change in the
equipment usually leads to a different focus in research, as
phenomena, which were previously out of range, can now be analysed
in a scientific manner.

In particular the advent of DFT and the continuous improvement of
computing algorithms applied to solid state physics have created an
environment in theory, where the inability to solve equations
analytically is no hindrance. Two of the main problems, needed to
solve the non-linear problem, are already incorporated in standard
DFT codes: (i) The problem of finding the charge density
$\rho_{i+1}$ after a given iteration $i$. This is done by elaborate
mixing schemes, which are constructed to speed up the convergency
process. (ii) The problem of finding the minimum energy density.
This problem is not different from the problem of finding it in
standard DFT convergency cycles, where one aims at processing with
the calculation along a hyper-surface, which will lead to a minimum.
In fact, if the charge density is a unique functional for a given
system, then the convergence criterium can be reformulated in terms
of the difference of charge in two subsequent iteration cycles: if
this difference is zero, then the charge density is the true
groundstate charge density. Thus the problem of finding the solution
of the non-linear eigenvalue problem and the groundstate charge
density from:
\begin{equation}
\left[ -2f \nabla^2 + v({\bf r}) \rho({\bf r}) \right]\Psi({\bf
r}) = \epsilon \Psi({\bf r})
\end{equation}
can be broken down in a number of discrete steps. Let us consider
a general case; a system with a number $M$ of ions, at the
positions ${\bf R}_i$ of our coordinate system. The number of
electrons $N$ shall be:
\begin{equation}
N = \int d^3{\bf r} \, \Psi^*({\bf r}) \Psi({\bf r}) = \int
d^3{\bf r} \,\rho({\bf r}) =  \sum_{i = 1}^M Z_i
\end{equation}
if the system is neutral and the ionic charge of atom $i$ is given
by $Z_i$. If we neglect for the time being magnetic systems, then
the potential $v({\bf r})$ is the sum of ionic and electronic
contributions; it can be written as (we use atomic units in the rest
of the paper):
\begin{equation}
v({\bf r}) = - \sum_{i = 1}^M \frac{Z_i}{|{\bf r} - {\bf R}_i|} +
\int d^3 {\bf r}' \frac{\rho({\bf r}')}{|{\bf r} - {\bf r}'|}
\end{equation}
Then the potential part of the eigenvalue equation is described by
(we follow the convention in DFT and write the product of external
potential and charge density as the effective potential
$v_{eff}({\bf r})$):
\begin{equation}
v_{eff}({\bf r}) = v({\bf r}) \rho({\bf r}) = - \sum_{i = 1}^M
\frac{\rho({\bf r}) Z_i}{|{\bf r} - {\bf R}_i|} + \int d^3 {\bf
r}' \frac{\rho({\bf r}) \rho({\bf r}')}{|{\bf r} - {\bf r}'|}
\end{equation}
The eigenvalue problem then has the form:
\begin{equation}
\left[ -2f \nabla^2 + v_{eff}({\bf r}) \right]\Psi({\bf r}) =
\epsilon \Psi({\bf r})
\end{equation}
A minimization process proceeds from one charge density to the next
without the calculation of Kohn-Sham eigenstates. One may wonder
that the parameter $f$, related to the kinetic energy density, still
contains the volume. This is a consequence of the rigorous
application of the energy principle, not only for the system as a
whole, but for every single point. If the density of charge is
increased, then the potential energy will also increase. This, in
turn means that the kinetic energy density will be lower. Given a
number of electrons in the system, all energy components depend on
the volume. The volume must therefore also show up in the kinetic
energy component.

\subsection{Initial charge density}

It is customary in DFT simulations to start a self-consistency cycle
with an initial guess about the charge density distribution. In
principle, this guess could involve any arbitrary choice, as long as
the total number of electrons in the system is equal to $N$.
However, most codes employ a superposition of atomic charges. As the
previous section and the zero order approximation of the non-linear
problem show, this is a suitable choice for the present problem.
Since the charge density will ultimately be computed
self-consistently, we may employ this approximation also for our
initial charge distribution, setting $\rho_0({\bf r})$ to:
\begin{equation}
\rho_0({\bf r}) = \sum_{i=1}^M \sum_{\alpha = 1}^{Z_i}
\rho_{\alpha,i}({\bf r} - {\bf R}_i)
\end{equation}
where $\alpha$ indicates the atomic orbitals of a given atom at
the position ${\bf R}_i$. From this initial charge density the
potential $v_{eff,0}({\bf r})$ is computed with:
\begin{equation}
v_{eff,0}({\bf r}) = - \sum_{i = 1}^M \frac{\rho_0({\bf r})
Z_i}{|{\bf r} - {\bf R}_i|} + \int d^3 {\bf r}' \frac{\rho_0({\bf
r}) \rho_0({\bf r}')}{|{\bf r} - {\bf r}'|}
\end{equation}

\subsection{Self consistency cycle}

The initial potential is the input for the first iterative cycle to
solve the ensuing {\em linear} density equation:
\begin{equation}
\left[ -2f \nabla^2 + v_{eff,0}({\bf r})\right] \Psi({\bf r}) =
\epsilon \Psi({\bf r})
\end{equation}
The problem is a standard problem in DFT, and a large number of
algorithms exists for its solution. Note that it is only necessary
to calculate the lowest eigenstate, defined by $\epsilon_1$, and
the corresponding real eigenvector $\rho_1({\bf r})$. This is in
marked contrast with today's methods, where an $N$ electron system
has to be solved for $N$ eigenvalues and eigenvectors. This means,
that the size of the problem depends only on the volume of the
system. If we consider that the volume usually scales linearly
with the number of atoms, then the ensuing numerical problem is
also of the same size. Apart from the simplification of not having
to solve the Schr\"odinger equation for single Kohn-Sham states
the new method scales also much better than usual codes, which
scale quadratically or cubic with $N$. It is a true order-$N$
method, or provides the optimum scaling with the size of the
system. Today, order-$N$ methods are based on a clever use of
density matrices (see e.g \cite{hg95}), or they use atomic
orbitals \cite{siesta}, which makes them unsuitable for metallic
systems. The suggested method has no limitation in this respect.
Once the many-electron wavefunction is known, the charge density
is given by:
\begin{equation}
\rho_1({\bf r}) = \Psi_0^*({\bf r}) \Psi_0({\bf r})
\end{equation}

The density $\rho_1({\bf r})$ is used to construct a new potential,
$v_{eff,1}({\bf r})$, by mixing the input density $\rho_0({\bf r})$
and the output density $\rho_1({\bf r})$ with the help of a suitable
mixing algorithm, so that:
\begin{equation}
\rho_1({\bf r}) = {\cal M}\left[\rho_0({\bf r}),\rho_1({\bf
r})\right]
\end{equation}
Again, this is standard procedure in today's DFT codes and a large
number of mixing schemes (see e.g.
\cite{pulay80,kerker80,broyden80}) exists to guarantee numerical
stability in the self-consistency cycle. The new potential
$v_{eff,1}({\bf r})$ is then:
\begin{equation}
v_{eff,1}({\bf r}) = - \sum_{i = 1}^M \frac{\rho_1({\bf r}) Z_i}{|{\bf
r} - {\bf R}_i|} + \int d^3 {\bf r}' \frac{\rho_1({\bf r})
\rho_1({\bf r}')}{|{\bf r} - {\bf r}'|}
\end{equation}
And the second iteration cycle consists again of solving a linear
eigenvalue problem described by:
\begin{equation}
\left[ - 2f \nabla^2 + v_{eff,1}({\bf r})\right] \Psi({\bf r}) =
\epsilon \Psi({\bf r})
\end{equation}
A solution of the problem leads to energy eigenvalue $\epsilon_2$
and eigenvector $\Psi_2({\bf r})$, which are used as an input for
the new effective potential $v_{eff,2}({\bf r})$. The iteration is
repeated until convergency, usually defined either by the change
of eigenvalues or the change of charge density, is achieved.

\subsection{Calculating the value of physical variables}

Once the groundstate charge density $\rho_g({\bf r})$ has been
computed the many-electron wavefunction can be determined by
solving the eigenvalue problem:
\begin{equation}
\left[- 2f \nabla^2 + v_g({\bf r}) \rho_g({\bf r})\right]
\Psi({\bf r}) = \epsilon \Psi({\bf r})
\end{equation}
The many-electron wavefunction may provide a direct way in the
future to compute all physical quantities of interest. Since the
many-electron wavefunction is the solution of the non-linear
Hamiltonian, the density matrix:
\begin{equation}
n({\bf r},{\bf r}') = \Psi({\bf r}) \Psi^*({\bf r}')
\end{equation}
can be used to calculate any physical quantity $A$ of interest via:
\begin{equation}
\langle A \rangle =  Tr \left[A \Psi({\bf r}) \Psi^*({\bf
r}')\right] = \int d^3{\bf r} A \rho({\bf r})
\end{equation}
However, the physical quantities like kinetic energy and potential
energy have been modified to obtain the general formulation.
Similarly, other quantities in the Schr\"odinger equation, e.g.
momenta, or electric field operators, will have to be adapted to the
non-linear formulation.

An alternative solution to the problem which allows to make use of
the vast body of theory developed, will be to use only the
many-electron groundstate charge density $\rho_g({\bf r})$, and to
construct the effective potential in the standard way by:
\begin{equation}
v_{eff,dft}({\bf r}) = v[\rho_g({\bf r})]
\end{equation}
This effective potential then allows to construct single particle
solutions in the standard manner:
\begin{equation}
\left\{- \frac{1}{2} \nabla^2 + v[\rho_g({\bf r})]\right\}
\psi_i({\bf r}) = \epsilon_i \psi_i({\bf r})
\end{equation}
These solutions can be used within the standard framework,
considering that the groundstate charge density completely
determines the physical properties of the system. The forces on ions
in a system can be calculated in the usual manner. Since the
Hamiltonian of the non-linear formulation is still hermitian, the
Hellman-Feynman theorem still applies \cite{feynman39}:
\begin{equation}
{\bf F}_{i} = - \frac{\partial E}{\partial {\bf R}_i} = - \int d^3
{\bf r} \rho({\bf r}) \frac{\partial v({\bf r})\rho({\bf
r})V}{\partial {\bf R}_i}
\end{equation}
From this relation the forces on atomic nuclei can be calculated in
the standard manner. These forces can then be used to optimize the
geometry of the system.

\subsection{Summary}

We have shown in this section that the non-linear formulation of
density functional theory allows to perform standard iteration
procedures as used for at least three decades. It was pointed out
that the main advantages of the new approach are:

\begin{enumerate}
\item  The iterations lead to the true many-electron charge
density. \item The complexity of the calculation scales linearly
with the number of atoms
\end{enumerate}

\section{Numerical tests}

Lang and Kohn showed in the early Seventies \cite{lang70,lang71}
that the electron charge at the surface of a metal oscillates as it
approaches the surface dipole. While standard DFT is able to capture
this behavior, the Friedel-oscillations, the Thomas-Fermi method
fails quite dramatically. It seems therefore an ideal test case for
non-linear DFT. The typical density of metals can be inferred from
the lattice type, the lattice constant, and the number of valence
electrons, assuming that the core covers only a small region of the
unit cell. The Wigner-Seitz radii of typical metals, as well as
their experimental workfunctions, are given in Table 1.
\begin{table}\label{table1}
\begin{center}
\begin{tabular}{l|c|c|c|c|c|r}
Metal & Type & a [a.u.] & V [a.u.$^3$] & e/cell &
$r_s$ & $\Phi$[eV] \\
  \hline
  Cs & bcc & 11.6 & 1560.9 & 2 & 5.74 & 1.81 \\
  Li & bcc & 6.63 & 291.4 & 2 & 3.26 & 2.38 \\
  Cu & fcc & 6.83 & 318.8 & 44 & 1.20 & 4.40 \\
  W & bcc & 5.98 & 214.0 & 12 & 1.62 & 4.50 \\
\end{tabular}
\caption{Lattice parameters, Wigner-Seitz radii of electrons, and
workfunction of typical metals. The experimental workfunctions were
taken from \cite{ashcroft}, the lattice parameters from
\cite{webelements}}
\end{center}
\end{table}

\subsection{Setup}

We performed self-consistent simulations of the groundstate charge
density for a metallic system, varying the density of the positive
background in the jellium calculation from $r_s$ = 3 to $r_s$ =
1.5a.u. The metal was modeled by a slab of 30($r_s$=1.5-2.0a.u.) to
40a.u.($r_s$ = 2.5-3.0a.u.) thickness, corresponding to more than 15
layers of the metal. In standard DFT simulations this is sufficient
to guarantee bulk properties in the center of the slab. In all
calculations the film was embedded in two vacua, the vacuum range
was 15-20 Bohr radii. This seems sufficient to mimic the surface
potential barrier, it should  also allow to estimate, whether the
density decays exponentially into the vacuum range, as found in
standard DFT simulations \cite{lang70,lang71}, or inversely
proportional with the distance from the jellium edge, as required by
the asymptotic decay due to image potentials.
\begin{figure}[th]
\begin{center}
\includegraphics[width=\columnwidth]{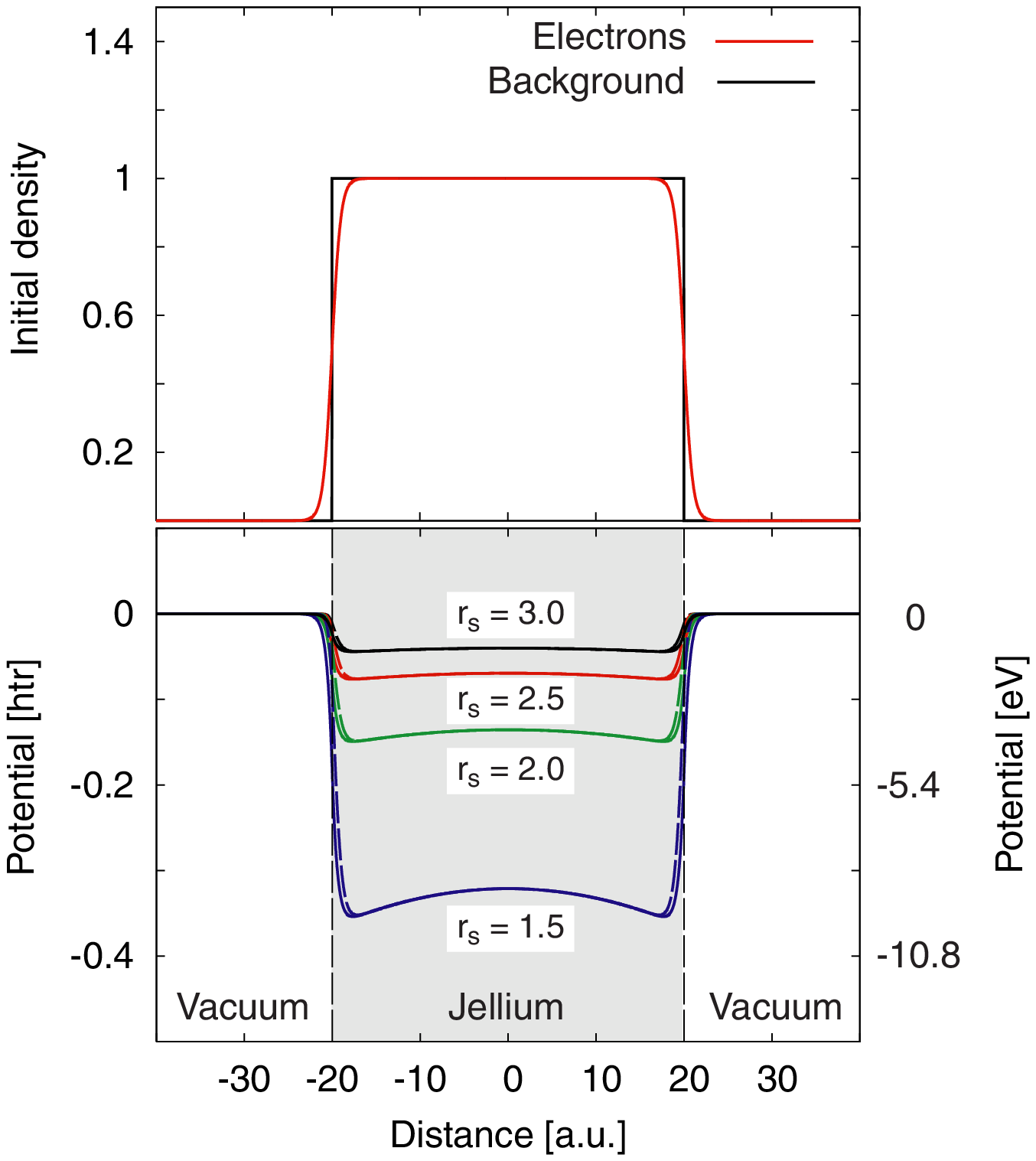}
\caption{(Color online) Setup for the calculation of metal surfaces.
The jellium film is embedded in two vacuum ranges (top). The initial
electrostatic potential (full graphs) inside the slab is equal to
the non-linear potential (dashed graphs) as defined in Eq.
\ref{nlpot_surf}. The potential varies with the density of the
background (bottom).} \label{fig021}
\end{center}
\end{figure}
The non-linear equation of the problem has the following form:
\begin{equation}
\left[-\frac{1}{2} \frac{d^2}{dz^2} + v_{nl}(z)\right] \Psi(z) = E
\Psi(z)
\end{equation}
Here, the eigenvalue $E$ is the energy value of the many-electron
electron charge. With a suitably chosen non-linear potential it
will be equal to the eigenvalue of the many-electron Schr\"odinger
equation. The dipole potential of the surface can be calculated in
the standard way by integrating the contributions from a point
deep inside the jellium into the vacuum
region\cite{lang70,lang71}:
\begin{equation}
v_{el}(z) = - 4 \pi \int_{z}^{\infty} dz' \int_{z'}^{\infty} dz''
\left[\rho(z'') - \rho_+(z'')\right],
\end{equation}
and by using the boundary condition $v_{el}(z \rightarrow \infty)$ =
0. The nonlinear potential was obtained by multiplying the dipole
potential with $V_{ws}\rho(z)$, where $V_{ws}$ is the Wigner-Seitz
volume:
\begin{equation}\label{nlpot_surf}
v_{nl}(z) = v_{el}(z) V_{ws}\rho(z) \qquad V_{ws} =
\frac{4\pi}{3}r_s^3
\end{equation}
The initial surface dipole was mimicked by a Fermi distibution
function at both jellium edges. The initial density in the vacuum at
a distance of 20a.u. was below 10$^{-17}$ in units of the jellium
density. The initial nonlinear  potential is nearly equal to the
electrostatic potential inside the slab, the only significant
differences occur at the jellium edges (see Fig. \ref{fig021}).
Conveniently, the nonlinear potential also has the same dimensions
as the electrostatic potential. It starts deviating only, once the
electron charge density in the initial setup differs substantially
from the positive background.
\begin{figure}[th]
\begin{center}
\includegraphics[width=\columnwidth]{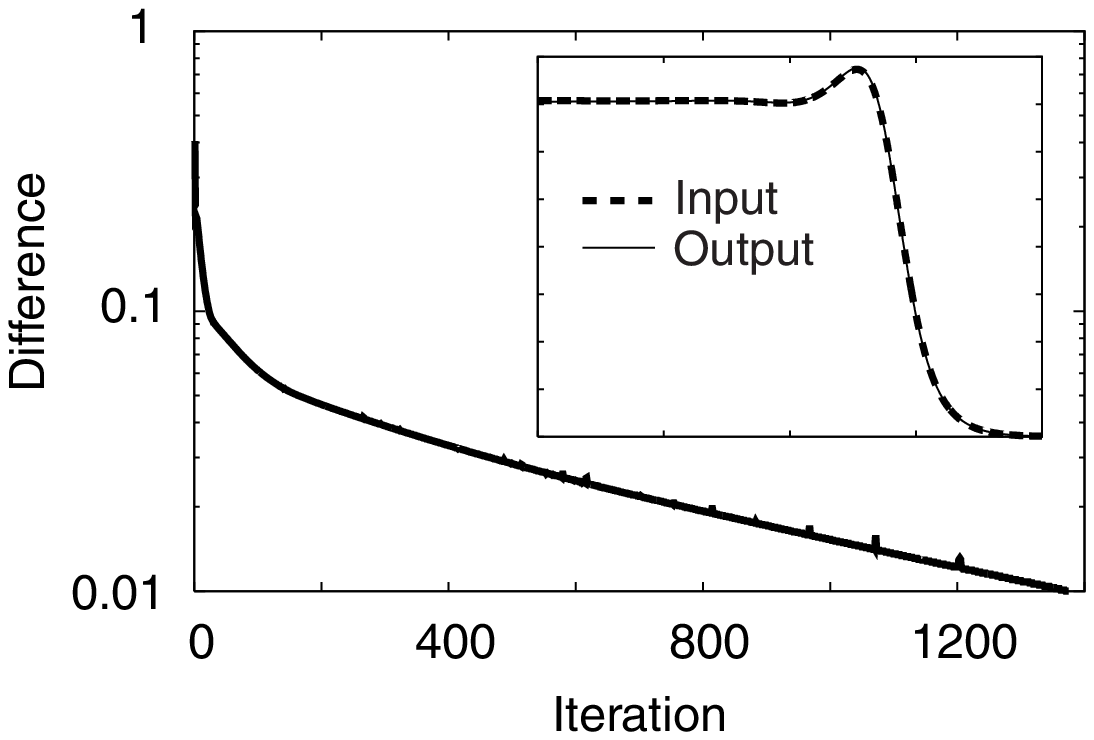}
\caption{Convergency of the iterations. The difference between input
and output charge density decreases close to exponentially to its
final value below 0.01. In the final iteration input and output
charge density are identical (inset).} \label{fig022}
\end{center}
\end{figure}

\subsection{Iterations}

The nonlinear equation has been solved by numerical integration. We
employed the Numerov integration scheme \cite{numerov}. The value of
$\Psi_{n+1}$ on an equally spaced z-grid is the convolution of the
preceding $\Psi$-values and the second derivative of the equations,
according to:
\begin{equation}
\Psi_{n+1} = \frac{2 \Psi_n - \Psi_{n-1} +
\frac{h^2}{12}\left(U_{n+1} + 10 F_n + F_{n-1}\right)}{1 + \frac{h^2
V_{n+1}}{12}}
\end{equation}
The precision of the result in one step is of the order $h^5$, which
renders the method quite accurate for one dimensional applications.
$h$ is the stepsize in the integration, the derivatives $F_n$ are
given by:
\begin{equation}
F_n = 2 \left(v_{nl,n} - E\right) \Psi_n
\end{equation}
In case of the non-linear equation $U_n$ will be zero at every
gridpoint, while $V_n$ is  equal to $2 \left(v_{nl,n} - E\right)$.
The integration is performed twice: once starting from the left,
integrating to the right, and once starting from the right and
integrating left. The two functions are matched at $z=0$. The
parameter $E$, the eigenvalue of the system, determines whether they
can actually be matched. The standard procedure in this case is to
compute the derivatives of the two functions, $\Psi_{left}$ and
$\Psi_{right}$ and to change the eigenvalue $E$ until the two
functions match. Numerical convergency with a simple linear mixing
scheme required a very high number of gridpoints (8000 gridpoints
over a length of 60-80a.u.) and very low mixing parameters (0.001 to
0.002) in the simulation. The difference between input and output
charge density is given by the integral:
\begin{equation}
N_d = \frac{1}{N} \int_{-\infty}^{+\infty} dz  |\rho_{in}(z) -
\rho_{out}(z)|
\end{equation}
The charge density in the iterations was converged to a value of
0.01, which seemed sufficiently precise. The eigenvalue at this
level of convergency is precise to about 1meV. The difference
between input and output charge density in the iterations decreases
exponentially, as shown in Fig. \ref{fig022}.

\subsection{Density oscillations}

\begin{figure}[th]
\begin{center}
\includegraphics[width=\columnwidth]{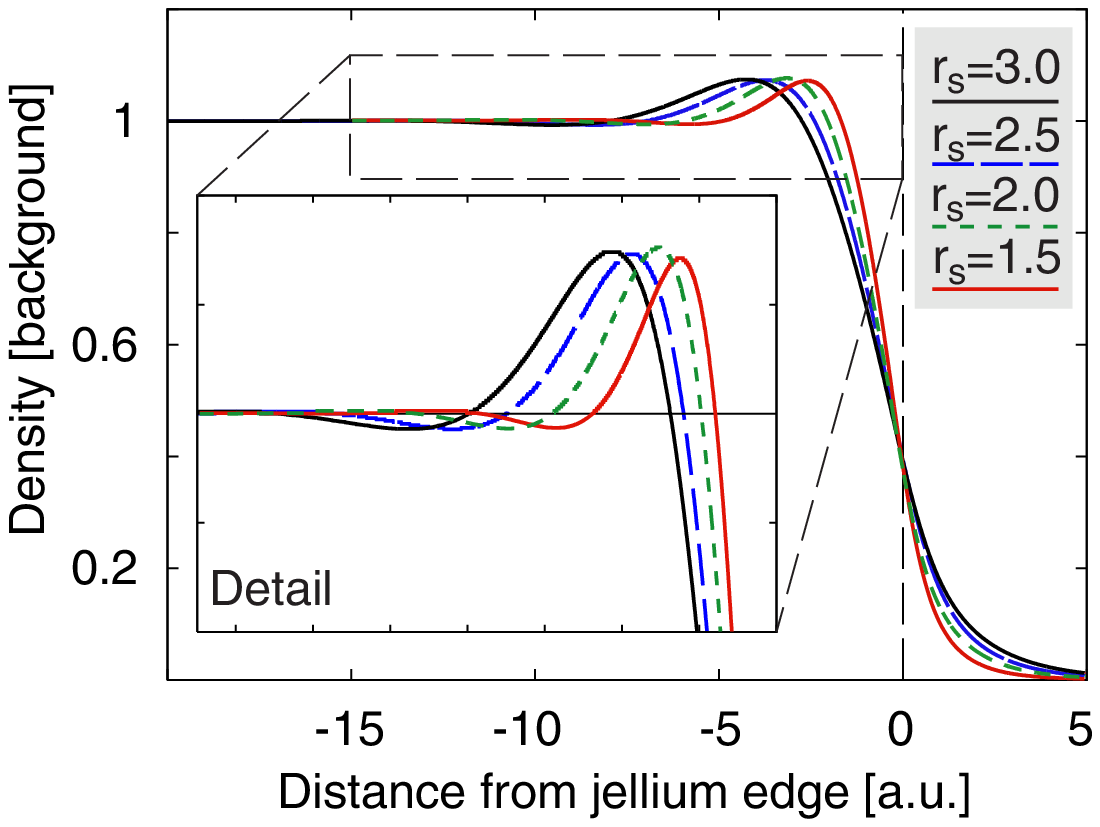}
\caption{(Color online) Density oscillations at the jellium edge.
The density shows a characteristic peak at the boundary. The maximum
value of the density is nearly equivalent in all simulations and
about 8\% higher than the density of positive charge. The
oscillations decay into the bulk, the characteristic wavelength
decreases with increasing density (see detail).} \label{fig023}
\end{center}
\end{figure}

In the simulations we find, in every case, an oscillation of the
charge density at the jellium edge. The wavelength of the
oscillation depends on the jellium density. Contrary to Lang and
Kohn \cite{lang70,lang71} we do not find that the oscillations are
damped for values of $r_s <$ 2.0a.u. The results of the simulations
are shown in Fig. \ref{fig023}.
We find the same increase of electron density at the jellium edge
for every density value. The surplus charge is about 8\% compared
to the bulk value. The result suggests that within the non-linear
framework a gradual removal of the peak and an approach of the
Thomas-Fermi result in the limit of $r_s \rightarrow
2$\cite{lang70} does not occur. The standard DFT result for a
uniform positive background cannot be extended beyond $r_s$=3, as
Lang and Kohn showed in their calculations \cite{lang70,lang71}.
This feature of the jellium model corresponds to a failure to
account for the experimental surface energies in this range. At
present, we have not yet developed a method to calculate surface
energies from the many-electron charge densities alone. However,
it seems quite possible that results remain reasonable also in
this range, considering that the surface dipole seems independent
of the background charge. Analyzing the details of the density
oscillations into the bulk we find a shortening of the wavelength,
as the density increases. This is mainly due to changes in the
non-linear potential: as the potential becomes more negative, the
eigenvalues do not keep up. The positive energy difference between
the energy of the electron charge and the potential increases in
this case, which leads to shorter wavelengths (see next sections).

\subsection{Potentials and eigenvalues}

\begin{figure}[th]
\begin{center}
\includegraphics[width=\columnwidth]{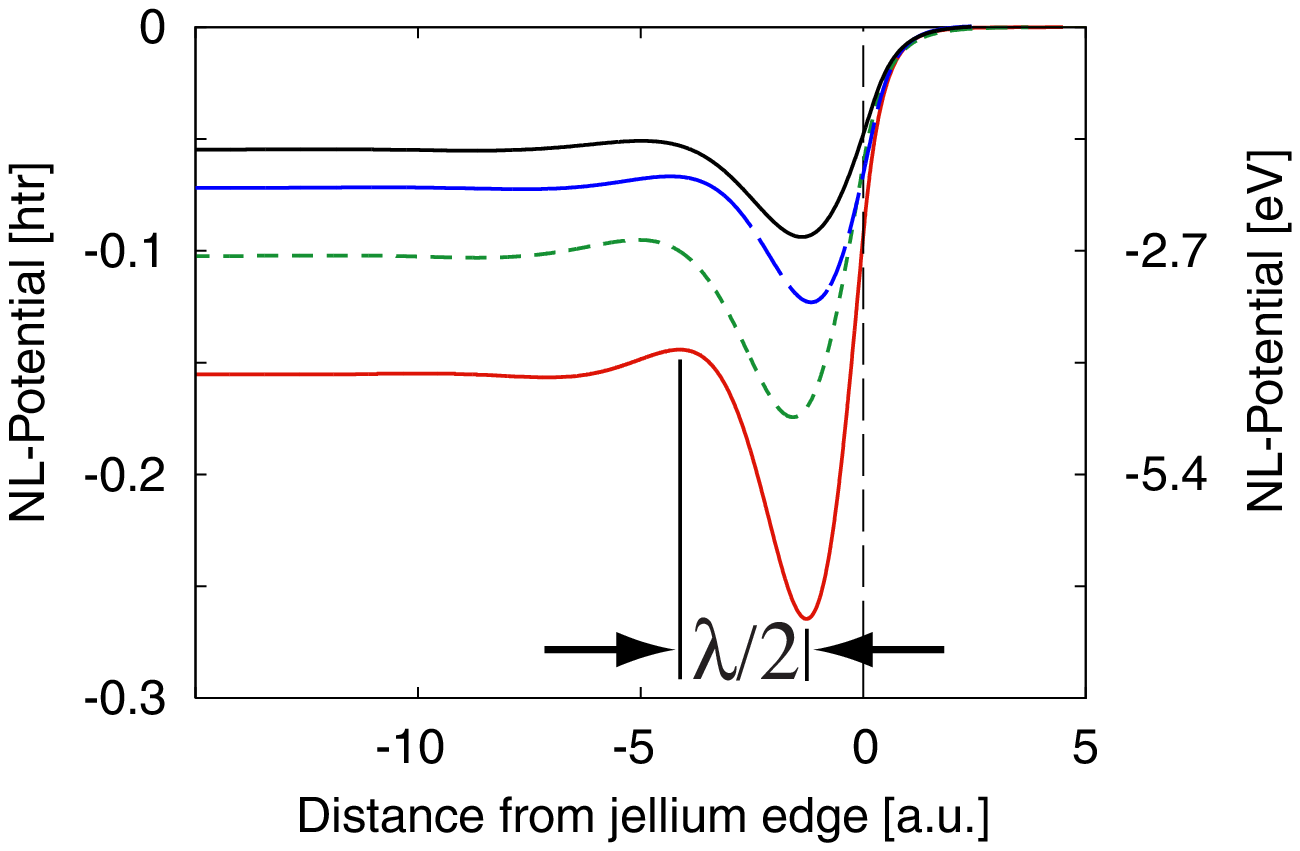}
\caption{(Color online) Non-linear potential of the fully converged
systems. The surface dipole corresponds to a minimum of the
potential near the jellium edge. The characteristic wavelength
$\lambda$ decreases with increasing density.} \label{fig024}
\end{center}
\end{figure}

The peak of the charge density at the jellium edge corresponds to
a trough in the attractive potential. Compared to the initial
potentials (see Fig. \ref{fig022}) the main changes are the
deepening of the dip at the barrier and the potential oscillations
decaying into the bulk (see Fig. \ref{fig024}). Analyzing the
actual process it has to be remembered that the many-electron
wavefunction will be in phase throughout the film. As the two
partial waves are matched at the center of the film, the actual
wavelength is not completely arbitrary but reflects also the
boundary conditions.
From a physical point of view this is a somewhat unrealistic
setup, because the waves will be reflected at the core of bulk
atoms. The ensuing multiple scattering effects should change the
distribution of maxima and minima in a way not reproducible in a
one-dimensional model. However, it is quite clear from the results
that the many-electron wavefunction will have to be in-phase over
a lengthscale of a few nanometer. The coherence can only be broken
by electron-phonon interactions in a thermally activated
environment. We would therefore conclude that long range effects
are a necessary consequence of the present model.

The eigenvalues obtained in the simulations are given in Table 2.
They agree very well with experimental data, considering that the
model is relatively simple. The only deviations are observed in the
low density range $r_s > 3.0$. This could be due either to
neglecting the core volume in the calculation of the Wigner-Seitz
radius, or to the fact that electron charge in a low density
environment will experience the point-like and regular arrangement
of the ionic cores much stronger than in a high density environment.
In the first case, the actual Wigner-Seitz radius is smaller than
the assumed value, in the second case the jellium model is no longer
suitable. In experiments the measured values are in the range of 2.0
to 5.0eV. This corresponds to a WS radius in the range of 2.5 to
1.3a.u. Comparing our results for the workfunction with full
potential DFT simulations \cite{hofer99}, we note that the agreement
with experiments for standard transition metals is improved.  While
in DFT simulations of 5$d$ metals the typical workfunctions are well
above 5eV, in case of tungsten even above 6eV, they are still in the
range of 4-5eV in the present model.  We may conclude that the
simulations reproduce the experimental values reasonably well, and
that the problems occurring in jellium simulations of high density
could be due to standard density functional theory itself.
Physically speaking, one would expect that a high density of
electron charge makes a continuous model more applicable than a low
density environment. In standard DFT jellium simulations, one
observes exactly the opposite trend \cite{lang70,lang71}.
\begin{table}\label{table2}
\begin{center}
\begin{tabular}{l|cccc}
$r_s$ [a.u.] & Eigenvalue [htr] & $\Phi$ [eV] & Metal & $\Phi_{exp} $[eV]\\
  \hline
  3.0 & -0.055 & 1.51 & Li & 2.38 \\
  2.5 & -0.073 & 1.98 & Mg & 3.64 \\
  2.0 & -0.102 & 2.80 & Al & 4.25 \\
  1.5 & -0.160 & 4.35 & W &  4.50 \\
\end{tabular}
\caption{Eigenvalues and workfunctions for different densities. The
workfunctions are in the range of 1.5 to 4.35eV, corresponding
approximately to the range observed in experiments. Experimental
values were taken from \cite{ashcroft}. The deviation in the low
density regime could be due to the neglect of the core region (see
the text).}
\end{center}
\end{table}

\subsection{Vacuum decay}

\begin{figure}[th]
\begin{center}
\includegraphics[width=\columnwidth]{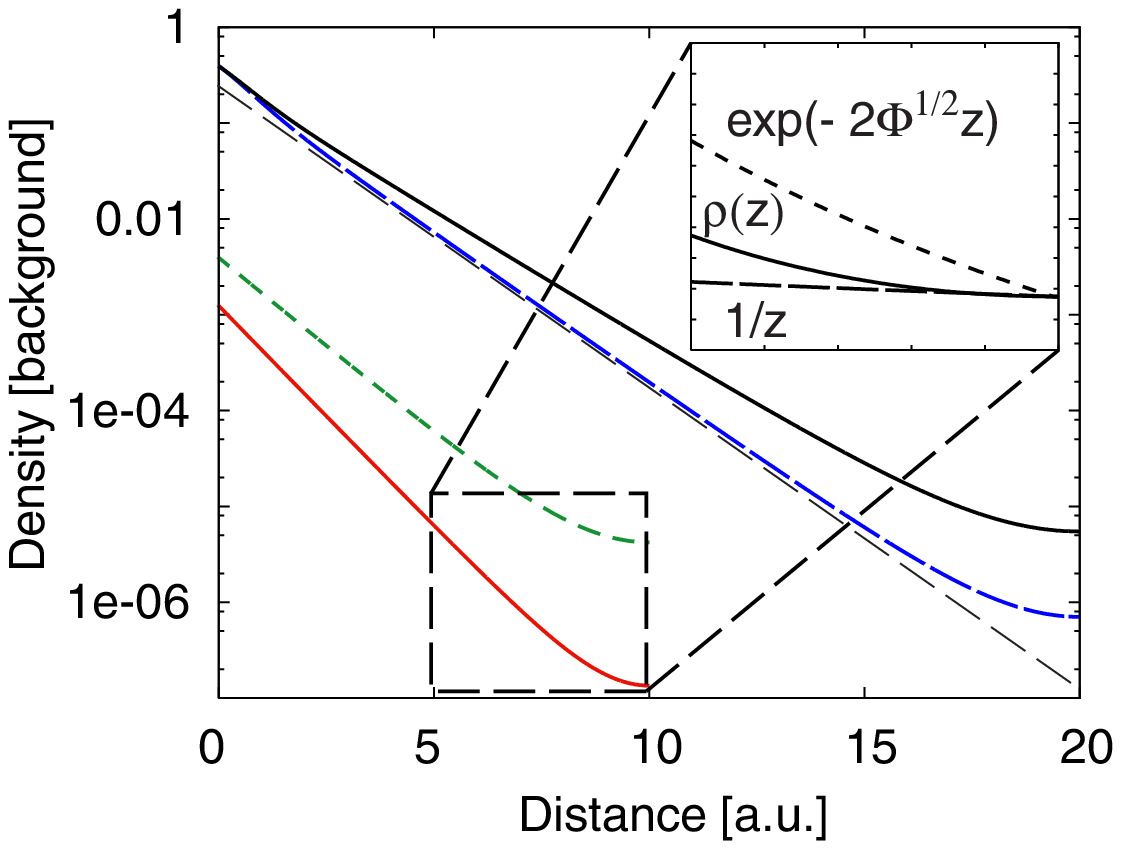}
\caption{(Color online) Density decay in the vacuum region. The
logarithmic plot shows a substantial deviation from an exponential
behavior in the high distance regime.} \label{fig026}
\end{center}
\end{figure}

Another reason for obtaining lower workfunctions than in standard
DFT could be the vacuum decay of electron density. The value for
$r_s$=3.0 at a distance of 20a.u. from the surface in the fully
converged system is about 10$^{-7}$e/a.u.$^3$. This is ten orders of
magnitude higher than at the beginning of the self consistency loop.
It is also three to four orders or magnitude higher than in standard
DFT simulations. A logarithmic plot of the vacuum density reveals
quite clearly, that the density {\em does not} decay exponentially
in the long distance limit. Logarithmic plots of all systems are
shown in Fig. \ref{fig026}. The curves deviate from an exponential
characteristic. The inset shows a fit of the density to two graphs:
(i) An exponential with the decay constant equal to $\sqrt{2\Phi}$,
and a function $1/z$. Both curves have the same value at the
ultimate limit of the calculation, 15a.u. from the jellium edge.
However, the exponential graph intersects the density curve at a
discrete angle, while the $1/z$ curve seems to match the asymptotic
density decay.
The match of the density to a $1/z$ curve is limited to a small
range of less than 5a.u. Even though the vacuum density is at least
three orders of magnitude higher than in standard DFT simulations,
the result is therefore not fully conclusive. We shall return to
this point in future publications.

\subsection{Kinetic energy functionals}

\begin{figure}[th]
\begin{center}
\includegraphics[width=\columnwidth]{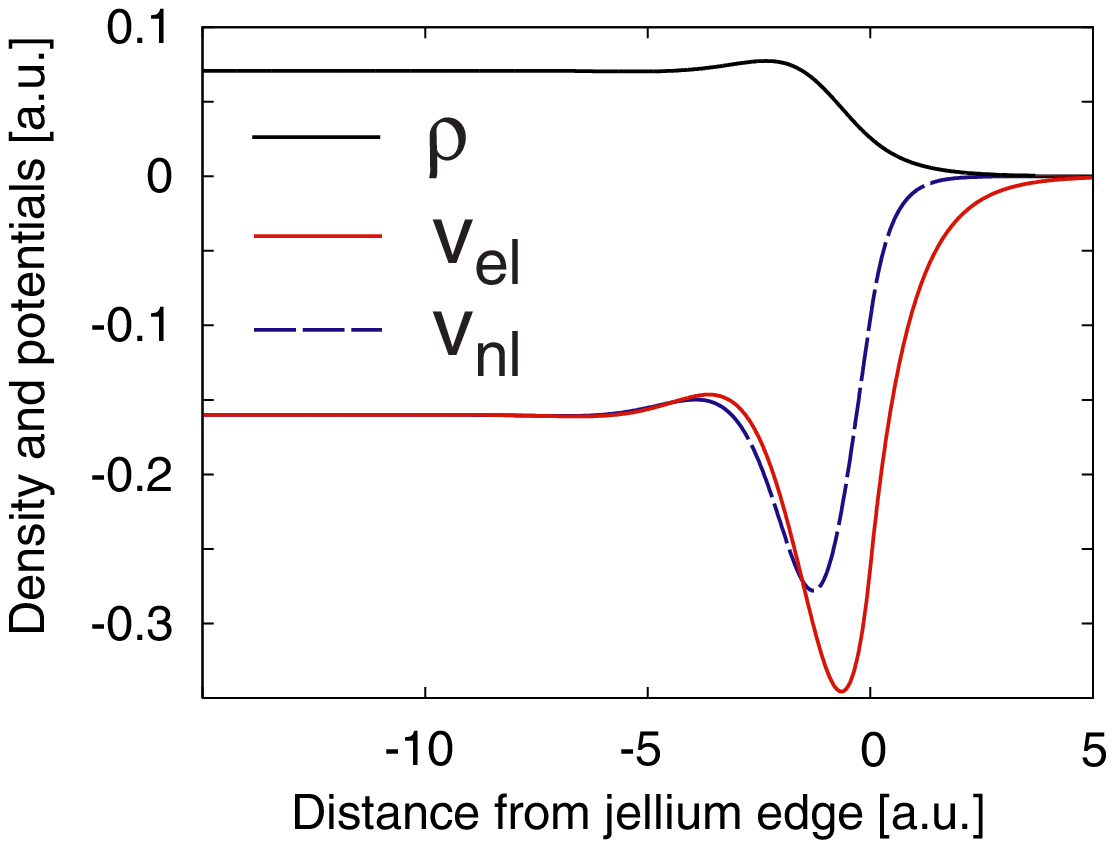}
\caption{(Color online) Density, electrostatic potential, and
nonlinear potential for $r_s = 1.5$a.u. The potential barrier around
-4a.u. screens surface electrons from bulk electrons.}
\label{fig027}
\end{center}
\end{figure}

The potentials, energy eigenvalues, and $k$-values for different
densities and two positions, at the center of the film, and at the
peak of the surface dipole, are given in Table 3. The density,
electrostatic potential, and nonlinear potential for $r_s = 1.5$a.u.
are shown in Fig. \ref{fig027}. The first conclusion drawn from the
numerical values is that the electrostatic potential is equal to the
nonlinear potential at the center of the slab. This is
understandable, if one considers that the density at this point is
the initial density; the product of Wigner-Seitz radius and density
is in this case equal to unity. Importantly, this means that there
is no correction to the electrostatic potential at this point. The
situation changes at the minimum of the surface dipole potential ($z
= z_0$), about 1.5a.u. from the jellium edge. Here, the charge
density deviates from the initial value, and in this case the
nonlinear potential will be affected. The difference, however is
small and less than 10\%, as the density peak does not exceed this
surplus value. Concerning the eigenvalues, we find in all cases that
they are equal to the potential at the center of the film. This
means, that the kinetic energy density at the center of the film
must be zero. A feature, which is only possible if the curvature of
the wavefunctions at this point vanishes. The density oscillations
at the jellium edge therefore decay into the bulk. Since the
potential at the surface dipole is lower than the eigenvalue, the
wavefunctions and also the densities will oscillate. Basing the
oscillations, and the relation between $k$-values and energies of a
free electron gas in a potential $v_{nl}(z_0)$, one would expect the
$k$-values to comply with the dispersion relation:
\begin{equation}
\frac{1}{2} k^2 = E - v_{nl}(z_0)
\end{equation}
As an inspection of the numerical data shows, this is not correct.
In fact, the term on the left is about five times higher than the
term on the right. This means, clearly, that the kinetic energy of
the nonlinear equation is not the usual single-particle or
Thomas-Fermi kinetic energy. The difference becomes clear if one
remembers that the fundamental change in the Schr\"odinger equation
was dividing the kinetic single-electron term by the volume of the
system. In a one-dimensional model, this corresponds to the
lengthscale of the system. Here, we have to consider that the
surface dipole is approximately 4-5a.u. wide: dividing the term on
the left by 5 leads to the correct relation. Physically speaking,
the change of the fundamental equation transformed the kinetic
single-electron term into a kinetic energy functional. The decisive
parameter of this functional is the confinement of the electrons in
the field of the surface dipole.
\begin{table}\label{table3}
\begin{center}
\begin{tabular}{l|ccccccc}
  \hline
  $r_s$ & $v_{el}(0)$ & $v_{el}(z_0)$ & $v_{nl}(0)$ & $v_{nl}(z_0)$ & $E$ &$E - v_{nl}(z_0)$ & $k$ \\
  \hline
1.5 & -0.160 & -0.290 & -0.160 & -0.265 & -0.160 & 0.105 & 1.080 \\
2.0 & -0.102 & -0.190 & -0.102 & -0.174 & -0.102 & 0.072 & 0.910
\\
2.5 & -0.073 & -0.136 & -0.073 & -0.124 & -0.727 & 0.051 & 0.760
\\
3.0 & -0.056 & -0.104 & -0.056 & -0.095 & -0.556 & 0.039 & 0.660
\\
\end{tabular}
\caption{Potentials at the center of the film $z = 0$, at the peak
of the surface dipole $z = z_0$, eigenvalue $E$, difference between
eigenvalue and attractive potential, and wavevector $k$. Note that
the $\frac{1}{2}k^2$ is about five times larger than the difference
between eigenvalue and attractive potential at $z_0$. The difference
bears on the confinement of electrons, included in the kinetic
energy functional (for details, see the text).}
\end{center}
\end{table}
Regarding the energy eigenvalues it is quite clear that the energy
density at the center of the slab is the same as the potential
density. Approaching the jellium edge, the potential changes and the
kinetic energy component changes accordingly: The system is indeed
characterized by a constant energy density throughout. It is,
therefore, a system in energetic equilibrium. The potential minimum
between -7a.u and -3a.u. leads to an exponential decay of the
wavefunction into the potential barrier and effectively screens the
surface dipole from the bulk charge. This feature contains already
the essence of a surface state, observed on many metal surfaces.

\subsection{Summary}

The method was tested for a metallic surface within a continuous
background model, varying the density over a range from $r_s$=1.5 to
$r_s$ = 3.0a.u. In this case we found, similarly to the results in
standard DFT, that the density oscillates at the jellium edge, and
that the wavelength of the oscillations depends on the jellium
density. We compared the calculated eigenvalues with experimental
values for the workfunctions and found good agreement, in particular
in the high density range. The kinetic energy functional in the
simulations includes the confinement of electron charge. This has
the effect that the energy density is constant throughout the
system. Finally, we determined the decay of the density in the
vacuum range. Here we find that the asymptotic behavior of the
vacuum charge seems to be better described by a $1/z$ characteristic
than an exponential function. In all simulations the only input
parameters were the background density and the thickness of the
slab.

\section*{Acknowledgements}

WAH is indebted to George Darling and Jacob Gavartin for stimulating
discussions, which to a large extent initiated or focussed research
in this particular direction. Both authors thank Andres Arnau for
sharing the jellium code, and Gilberto Teobaldi for discussions and
corrections.  KP is funded by the European Commission under project
NMP3-CT-2004-001561. WAH thanks the Royal Society for the award of a
University Research Fellowship, providing the necessary time to
pursue this investigation.


\begin{thebibliography}{99}
\bibitem{hk64}
S. Hohenberg and W. Kohn, Phys. Rev. 136, B864 (1964)
\bibitem{ks65}
W. Kohn and L. J. Sham, Phys. Rev. 140, A1133 (1965)
\bibitem{cp85}
R. Car and M. Parinello, Phys. Rev. Lett. 55, 2471 (1985)
\bibitem{hg95}
E. Hernandez, M. J. Gillan, and C. M. Goringe, Phys. Rev. B 53, 7147
(1996)
\bibitem{wl95}
W. Yang and T.-S. Lee, J. Chem. Phys. 103, 5674 (1995)
\bibitem{bhg97}
R. Baer and M. Head-Gordon, Phys. Rev. Lett. 79, 3962 (1997)
\bibitem{levy84}
M. Levy, J.P. Perdew, and V. Sahni, Phys. Rev. A 30, 2745 (1984)
\bibitem{dreizler90}
R. M. Dreizler and E. K. U. Gross, Density Functional Theory,
Springer, Berlin (1990) pp.62-64
\bibitem{wang99}
Y. A. Wang, N. Govind, and E. A. Carter, Phys. Rev. B 60, 16350
(1999)
\bibitem{hofer03}
W. A. Hofer, A. S. Foster, and A. L. Shluger, Rev. Mod. Phys. 75,
1287 (2003)
\bibitem{palotas05}
K. Palotas and W. A. Hofer, J. Phys.: Cond. Mat. 17, 2705 (2005)
\bibitem{datta95}
S. Datta, Electronic Transport in Mesoscopic Systems, Cambridge
University Press, Cambridge (1995)
\bibitem{bao03}
W. Bao, N. J. Mauser, and H. P. Stimming, Comm. Math. Sci. 1,
909-828 (2003)
\bibitem{ldb25}
L. de Broglie, Ann. Phys. 3, 22 (1925)
\bibitem{dg27}
C. Davisson and L. H. Germer Phys. Rev. 30, 705 (1927)
\bibitem{selleri88}
E. Selleri (ed.), Quantum Mechanics versus Local Realism, Plenum
Press, New York (1988)
\bibitem{schrodinger35}
E. Schr\"odinger, Naturwissenschaften 23, 807-812; 823-828; 843-848
(1935)
\bibitem{epr35}
A. Einstein, N. Rosen, and B. Podolsky Phys. Rev. 47, 180 (1935)
\bibitem{siesta}
P. Ordej\'on, D. A. Drabold, M. P. Grumbach and R. M. Martin, Phys.
Rev. B {\bf 48}, 14646 (1993); Phys. Rev. B 51, 1456 (1995)
\bibitem{pulay80}
P. Pulay, Chem. Phys. Lett. 73, 393 (1980)
\bibitem{kerker80}
G.P. Kerker, J. Phys. C.: Solid St. Phys, L189 (1980)
\bibitem{broyden80}
D. D. Johnson, Phys. Rev. B 38, 12807 (1988), and references therein
\bibitem{feynman39}
R. P. Feynman, Phys. Rev. 56, 340 (1939)
\bibitem{ashcroft}
N. W. Ashcroft and N. D. Mermin, Solid State Physics, Thmoson
Learning (1976) p. 364
\bibitem{webelements}
Cristallographic data were taken from http://www.webelements.com.
\bibitem{lang70}
N. D. Lang and W. Kohn, Phys. Rev B 1, 4555 (1970)
\bibitem{lang71}
N. D. Lang and W. Kohn, Phys. Rev. B 3, 1215 (1971)
\bibitem{numerov}
J. L. M. Quiroz Gonzaleza and D. Thompson, Computers in Physics {\bf
11}, 514 (1997)
\bibitem{hofer99}
W. A. Hofer, PhD thesis, Technische Universit\"at Wien (1999)
\end{thebibliography}
\end{document}